\documentclass[a4paper,fontsize=12pt]{scrartcl}

\usepackage[a4paper, total={6.5in, 7.5in}]{geometry}
\usepackage{xspace}
\usepackage{epsfig}
\usepackage{xcolor}
\usepackage{syntax}
\usepackage[fleqn]{amsmath}
\usepackage{amssymb}
\usepackage{semantic}
\usepackage{hyperref}
\usepackage{lmodern}
\usepackage[T1]{fontenc}
\usepackage{makeidx}
\usepackage{float}
\usepackage{alltt}

\makeindex
\title{\LARGE{QRscript specification}\\
{\Large (Version 0.9)}}
\author{}

\begin{document}
\maketitle

\begin{abstract}
This document \cite{QRscript-spec} reports the specifications of QRscript, which is a series of rules on how to embed a programming language into a QR code. 

Authors of this specification document are:
\begin{itemize}
    \item Stefano Scanzio (CNR-IEIIT, \href{https://www.skenz.it/ss}{https://www.skenz.it/ss}, stefano.scanzio[at]cnr.it)
    \item Matteo Rosani (matteo.rosani[at]gmail.com)
    \item Mattia Scamuzzi (mattia.scamuzzi[at]gmail.com)
    \item Gianluca Cena (CNR-IEIIT, gianluca.cena[at]cnr.it)
\end{itemize}

\end{abstract}

\clearpage
\tableofcontents

\newpage

\pagestyle{myheadings}
\markright{QRscript specification (Version 0.9)}

\section{Scope}
This specification document specifies the syntax and semantics of QRscript. The current document only shows the part related to the QRscript header, i.e., the first part of the binary code that must be inserted into the QR code. A QR code containing an executable code is called an executable QR code (eQR code). QRscript supports different dialects, i.e., sublanguages with implementation characteristics specific to the application field. The specifications of the individual dialects will be described in separate documents.

\section{Conformance}
A completely conforming implementation of QRscript must provide and support all the functionalities mandated in this specification. The word \textit{must} highlights these mandatory specifications. Some specific names that are important in this specification were highlighted in \textit{italic}.

\section{Normative References}
The following referenced documents are indispensable for the application of this document.
For dated references, only the edition cited applies. For undated references, the latest edition of the referenced document (including any amendments) applies.

\begin{enumerate}
\item The QR code standard, ISO/IEC 18004:2015 \cite{QRstandard}
\item The UTF-8 standard, RFC 3629 \cite{rfc3629}
\end{enumerate}

\section{Overview}
The possibility of inserting a high-level programming language inside a QR code to obtain an eQR code was first presented in \cite{9921530}, which defined the main ideas of QRscript that are expanded and formalized in this specification document.

\begin{figure}[H]
	\begin{center}
	\includegraphics[width=0.8\columnwidth]{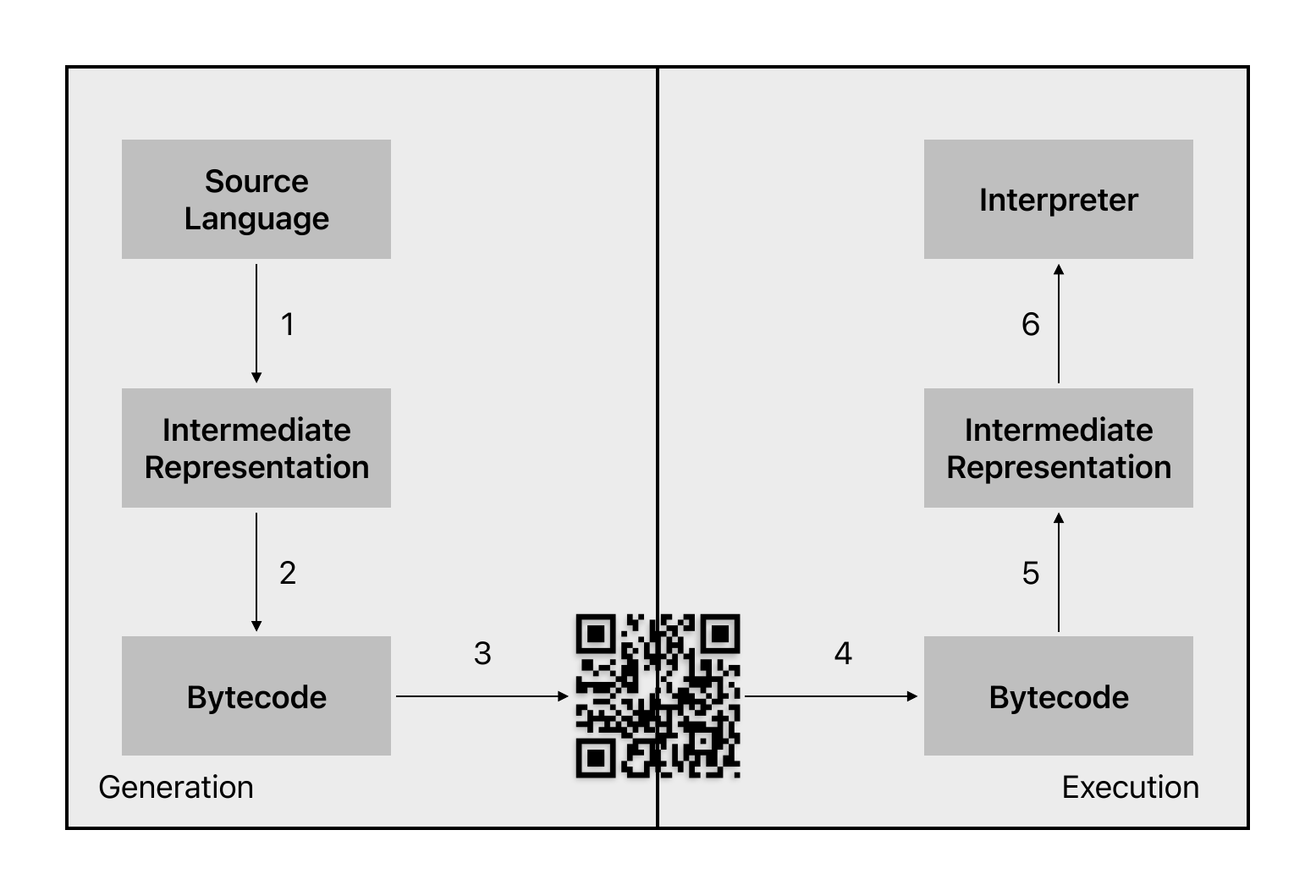}
	\end{center}
	\caption{Chain of usage of eQR code technology (generation and execution processes)}
	\label{fig:process}
\end{figure}

The whole chain of usage of this technology can be schematized as shown in Fig.~\ref{fig:process} in which the process is divided into a \textit{generation} phase and an \textit{execution} phase.

The generation phase leads to the creation of the eQR code, and it is detailed below:
\begin{enumerate}
    \item A \textit{high-level language} (Source Language) is translated into an \textit{intermediate representation} (Intermediate Representation). It does not matter which high-level language is used as it is not the topic of this specification document. Different high-level languages can be translated to the same intermediate representation.
    \item The intermediate representation is translated to a \textit{binary representation} (Bytecode) named \textit{eQRbytecode}.
    \item The eQRbytecode is transformed into an \textit{eQR code} containing the bytecode.
\end{enumerate}

The execution phase leads to the reading of the eQR code and its usage, and it is detailed below:
\begin{enumerate}
    \setcounter{enumi}{3}
    \item The eQR code is read and transformed in eQRbytecode.
    \item The eQRbytecode is converted into an intermediate representation.
    \item The intermediate representation is executed inside of a \textit{virtual machine} on board of the \textit{end-user device}. An end-user device is any apparatus capable of executing the virtual machine (i.e., a smartphone, a PC, an embedded device, etc.)
\end{enumerate}

The specifications of QRscript refer to the definition of the binary code (i.e., the eQRbytecode) that has to be saved inside of the eQR code and the rules used for its codification/decodification. In particular, these specifications represent partially the arrows 2 and 5 of Fig.~\ref{fig:process}. Some hints regarding arrows 1, 3, 4, 6 are provided just to make this description clearer.

In addition, this specification document only refers to the header of QRscript, the first part of the binary eQRbytecode that has to be inserted inside the eQR code. QRscript supports many dialects, i.e. sub-languages with implementation characteristics specific to the particular application scope (e.g., decision trees). The specifications of the different dialects will be described in separate specification documents.

Section~\ref{sec:eQRbytecodeHeader} will describe the characteristics of the eQRbytecode and, in particular, of the portion named \textit{QRscript header}, independent from the dialect, while Section~\ref{sec:QRcode} will report the characteristics in terms of typology of generated QR/eQR code and how the eQRbytecode is encoded within them.

\section{eQRbytecode: QRscript header}
\label{sec:eQRbytecodeHeader}
\begin{figure}[H]
	\begin{center}
	\includegraphics[width=0.8\columnwidth]{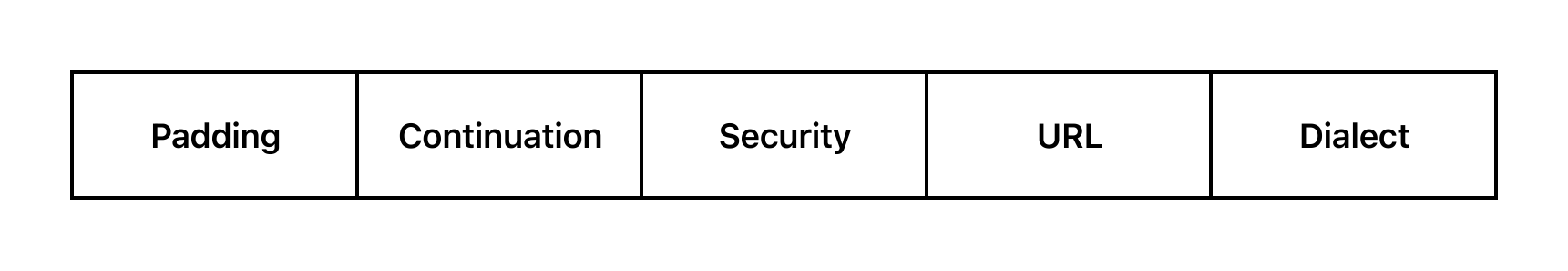}
	\end{center}
	\caption{QRscript header}
	\label{fig:header}
\end{figure}

Fig.~\ref{fig:header} shows the QRscript header, the first part of QRscript, which is divided into five sections: \textit{padding} (Subsection~\ref{sec:padding}), \textit{continuation} (Subsection~\ref{sec:continuation}), \textit{security} (Subsection~\ref{sec:security}), \textit{URL} (Subsection~\ref{sec:URL}), and \textit{dialect} (Subsection~\ref{sec:dialect}).

The \textit{padding} is used to pad the eQRbytecode to be written in the eQR code, in the case the selected coding is \textit{binary mode} and the number of bits of which it is composed is not a multiple of 8.

The \textit{continuation} is needed to split the program over multiple eQR codes in the event that the program is not small enough to fit in a single eQR code. The \textit{continuation} is optional, meaning that a compliant implementation does not have to implement it, but can directly set it as disabled by setting a \texttt{0} bit for the part of the QRscript header related to continuation.

The part about \textit{security} is the portion of the header of an eQR code that manages \textit{authenticity} and \textit{integrity}, and eventually the \textit{encryption} of the content of the eQR code.
The \textit{security} is optional, meaning that a compliant implementation does not have to implement it, but can directly set it as disabled by setting a \texttt{0000} bits for the part of the QRscript header related to security.

The \textit{URL} is used to permit the connection to an external URL (if an Internet or a local connection is available) and executes the application included in the external URL instead of that contained within the eQR code. This allows the execution of applications with more (interactive) content (such as video or images) than those that can be included within an eQR code. The \textit{URL} is optional, meaning that a compliant implementation does not have to implement it but can directly set it as disabled by setting a \texttt{0} bit for the part of the QRscript header related to the URL.

The \textit{dialect} section identifies the dialect and, therefore, the rules to be used for interpreting the eQRbytecode that follows the QRscript header. The QRscript header is followed by the \textit{dialect code}. It represents the program encoded in the eQR code.

\subsection{Padding}
\label{sec:padding}
\begin{figure}[H]
	\begin{center}
	\includegraphics[width=0.8\columnwidth]{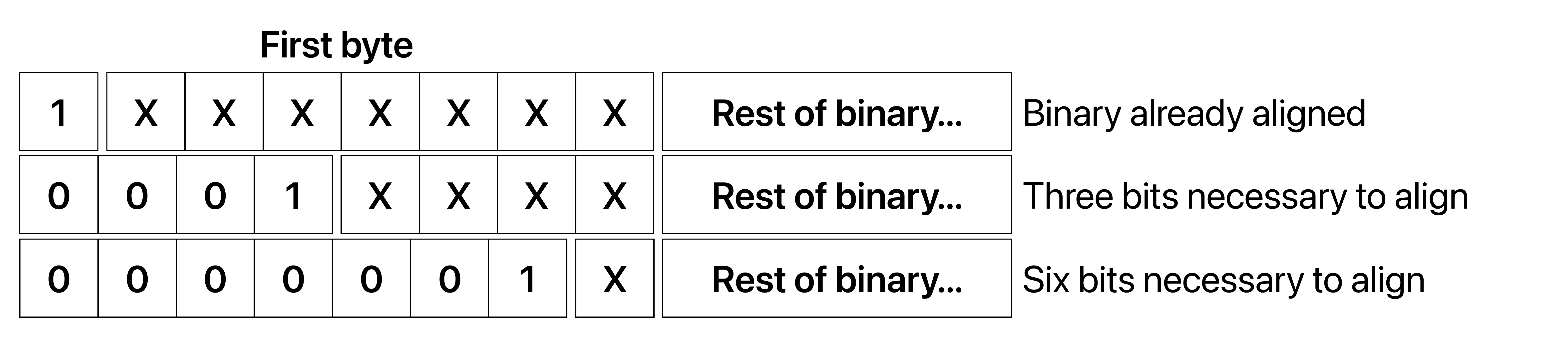}
	\end{center}
	\caption{Padding section}
	\label{fig:headerPadding}
\end{figure}
The QRbytecode header starts with a bit that indicates if there is any \textit{padding}, which is only required for the binary coding when the QRbytecode dimension is not divisible by 8 bits. Unlike communication protocols, in this context, the padding is located in the header (and not in the tail) because it should not depend on the type of dialect that is encoded in the eQR code. Specifically, the eQR code starts with \texttt{1} in the case of no padding, \texttt{01} means 1 bit of padding, \texttt{001} means 2 bits of padding, \texttt{0001} means 3 bits of padding, and so on. For numeric coding, the padding does not affect the conversion because the 0s before the first 1 are not significant. An example of padding is shown in Fig.~\ref{fig:headerPadding}.

\subsection{Continuation}
\label{sec:continuation}
\begin{figure}[H]
	\begin{center}
	\includegraphics[width=0.8\columnwidth]{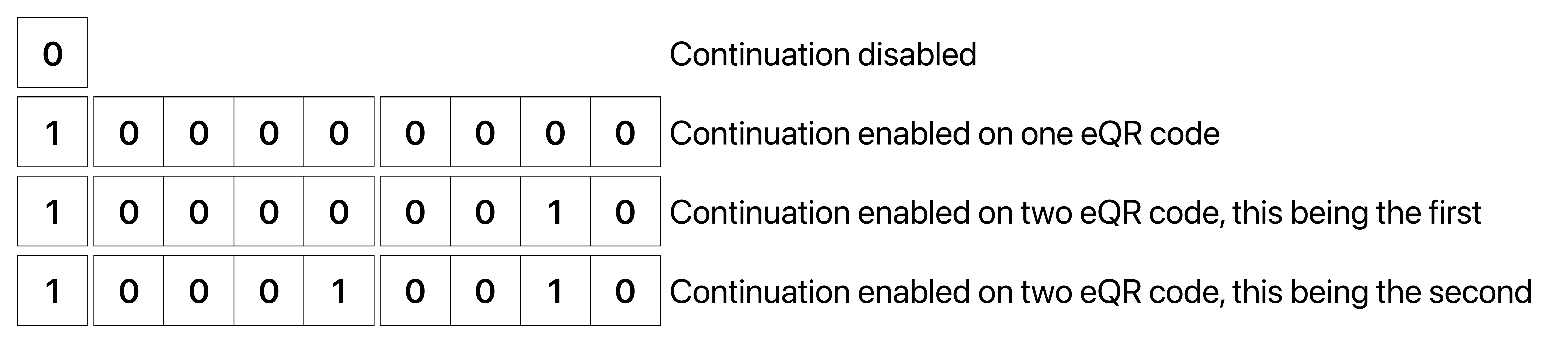}
	\end{center}
	\caption{Continuation section}
	\label{fig:headerContinuation}
\end{figure}
The size of a QR code is limited (for example, in version 40 with "low" error correction, the maximum amount of data that can be stored in a single QR code is 2953 bytes).

In order to overcome this limitation, QRscript provides the possibility to concatenate binary codes included in multiple eQR codes that are individually scanned by the device (called \textit{fragments}) into a single eQRbytecode. This is called \textit{continuation}.

The \textit{continuation} is enabled if the first bit of the eQRbytecode after the padding is set to \texttt{1}. This bit is then followed by the \textit{sequence number} assigned to that specific eQR code and the \textit{total number} of segments, which is the total number of segments that make up the program minus one. The \textit{sequence number} and \textit{total number} are encoded on $4$ bits, but they can be extended using the exponential encoding defined in Appendix~\ref{app:A}. During the reconstruction of the eQRbytecode, the application that performs the reading will use the sequence numbers to concatenate the binary codes contained in the various eQR codes in the correct order. The total number can be used by the application to communicate to the user which eQR codes still need to be scanned in order to rebuild the entire program.

An example of the encoding of a continuation is shown in Fig.~\ref{fig:headerContinuation}.

In an application that does not implement continuation and therefore does not allow the fragmentation of the application on multiple eQR codes, the first bits of the eQRbytecode can be \texttt{0} or \texttt{100000000} (i.e. a continuation composed of a single fragment).

\subsection{Security}
\label{sec:security}
\begin{figure}[H]
	\begin{center}
	\includegraphics[width=0.75\columnwidth]{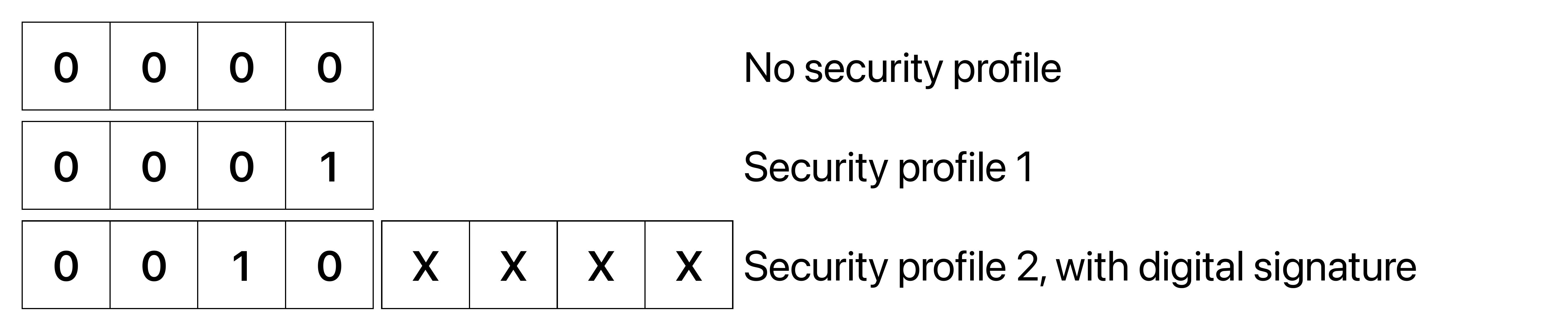}
	\end{center}
	\caption{Security section}
	\label{fig:headerSecurity}
\end{figure}
The \textit{security} is used to handle the \textit{authenticity}, the \textit{integrity}, and possibly the \textit{encryption}. It is optional, so a compliant implementation does not have to provide it. After the \textit{continuation}, the following 4 bits represent the security \textit{profile} used within the eQR code. This field follows the exponential encoding described in Appendix~\ref{app:A}. The sequence \texttt{0000} represents the absence of \textit{security} mechanisms. The other sequences represent specific \textit{security profiles}. The first $12$ security profiles are reserved and will be defined in subsequent versions of this specification document, while others can be used and defined by the application developers without any limitation.

A \textit{security profile} indicates:
\begin{enumerate}
\item which types of \textit{security} features are enabled, i.e., \textit{authenticity/integrity} and/or \textit{encryption}.
\item the parts of the eQR code that are involved. For example, although it is useful to provide \textit{authenticity/integrity} to the entire eQR code, it would be better to leave the \textit{continuation} part in clear in the case of \textit{encryption} to be still able to understand how to concatenate the various eQR codes without having to decrypt them.
\item The algorithm used to handle \textit{security}, e.g., RSA.
\item The \textit{length} of the digital signature (which may not be present) used to handle \textit{authenticity/integrity}, in bits. This data is very important because it is used for interpreting the subsequent bits as a digital signature and aligning correctly with the next part of the eQRbytecode.
\end{enumerate}

In the case of \textit{security}, the application developer will have to define a key-value pair, not formalized by this specification document, which reports a mapping between the bits representing the \textit{profile} (key) and the various defined profiles (value). Since this mapping is not defined by the specification document, different applications may have different key-value pairs. Remember that the first $12$ security profiles are reserved.

As anticipated, the subsequent bits can optionally represent the digital signature that can be used to manage \textit{authenticity} and \textit{integrity}.

\subsection{URL}
\label{sec:URL}
\begin{figure}[H]
	\begin{center}
	\includegraphics[width=0.8\columnwidth]{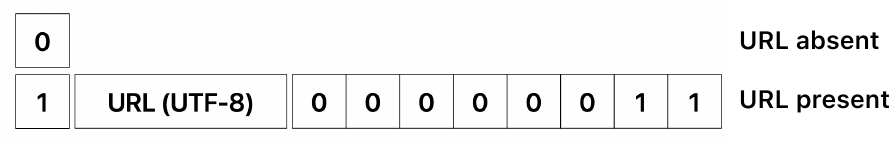}
	\end{center}
	\caption{URL section}
	\label{fig:headerURL}
\end{figure}
When the device of the final user can access the Internet, or at least a local connection, running a remote application from an \textit{URL} might be more suitable in some situations. For instance, it could enhance user interaction. The application runs on a remote web server, so it can include various information, such as videos and images, that are hard to incorporate within an eQR code due to space limitations. Therefore, an eQR code can contain a URL in the QRscript header. Specifically, a bit equal to \texttt{0} indicates that the URL is not present; otherwise, a bit equal to \texttt{1} identifies the presence of the URL. If the bit is \texttt{1}, the following bits encode the URL using UTF-8 encoding \cite{rfc3629}. The chosen character for the end of the string is \textit{end-of-text (EXT)} composed of the following bits \texttt{00000011}. The application running on end devices can program a particular response after requesting the selected URL, which refers back to the execution of the eQR code, allowing the possibility of deactivating the URL even after its creation. As previously reported, the \textit{URL} is optional, meaning that a compliant implementation does not have to implement it.

\subsection {Dialect}
\label{sec:dialect}
\begin{figure}[H]
	\begin{center}
	\includegraphics[width=1\columnwidth]{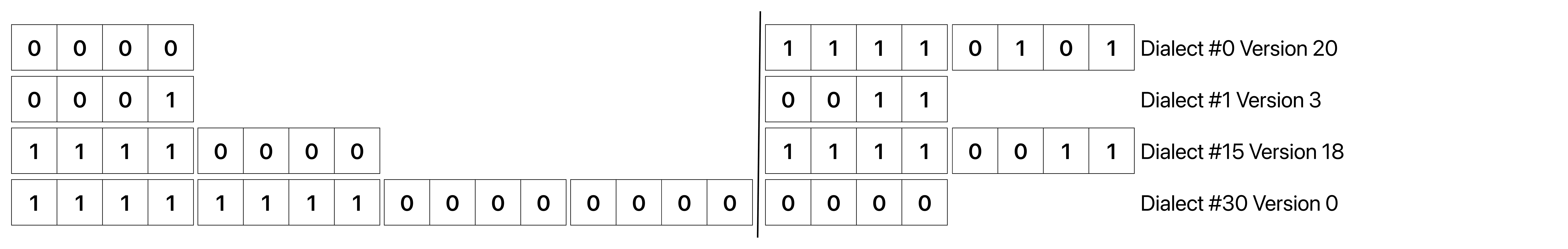}
	\end{center}
	\caption{Dialect section}
	\label{fig:headerDialect}
\end{figure}

After the URL, there are other $4$ bits that are used to identify the dialect that is being used in the following part of the eQR code, and other $4$ bits are aimed at identifying the version of the dialect. The dialect/version pair indicates which rules must be used to interpret the dialect that follows the QRscript header.
 
As previously mentioned, dialects are just languages with different capabilities in terms of the available instruction set and, consequently, with different sizes of the generated eQRbytecode. 

The work published in \cite{9921530} was, as far as we know, the first to present a programming language compiled and fit into a QR code. 
Nevertheless, in case some pre-existing languages were defined, they could easily be included in the proposed representation by assigning them a specific dialect code. 
The format for representing dialects and their versions is defined in such a way that it can be easily extended to a number of dialects/versions that exceed what can be encoded on four bits by exploiting the exponential encoding described in Appendix~\ref{app:A}.

At the time of writing, the available dialects are:
\begin{itemize}
    \item QRtree dialect, first four bits equal to \texttt{0000} (for the coding of decision trees) \cite{QRtree}
\end{itemize}

\section{eQR code coding}
\label{sec:QRcode}

QR codes are categorized in \textit{versions} (1 through 40), each with increasing size and capacity.

The encoding of data inside a QR code can be done in different \textit{input modes} with different information densities. Also, inside QR codes, an error detection and correction code is used that allows reading the content of partially damaged QR codes. How much of the QR code's data can be recovered is based on the \textit{correction level} chosen at the moment of creation.

The different \textit{input modes} are:
\begin{itemize}
    \item \textit{numerical}, stores only numerical characters (0-9) up to 7089 characters
    \item \textit{alphanumerical}, stores only alphanumerical characters (0-9, A-Z, space, `\$', `\%', `*', `+', `-', `.', `/', `:') up to 4296 characters
    \item \textit{binary}, stores raw binary data up to 2953 bytes
    \item \textit{kanji/kana}, stores characters of the Japanese alphabet up to 1817 characters
\end{itemize}

The different \textit{correction levels} are:
\begin{itemize}
    \item \textit{L level}, allows to recover around 7\% of data
    \item \textit{M level}, allows to recover around 15\% of data
    \item \textit{Q level}, allows to recover around 25\% of data
    \item \textit{H level}, allows to recover around 30\% of data
\end{itemize}

Based on previous information, the following design choices have been made.
In particular, the generated eQRbytecode must be encoded within the QR code using the \textit{binary} or the \textit{numeric} input mode. In the case of the \textit{binary} input mode, padding in the header is needed following the rules reported in Subsection~\ref{sec:padding} of this specification document.

\section{Main terms and acronyms}

\begin{itemize}

    \item \textit{Continuation}: feature of the QRscript header needed to split the program over multiple eQR codes.
    
    \item \textit{Dialect}: sub-languages with implementation characteristics specific to the particular application scope (e.g., decision trees, etc.).
   
    \item \textit{Dialect code}: program coded within the eQR codes, which specification is reported in separate documents.
    
    \item \textit{eQRbytecode}: the binary representation of the executable program. It is used to create an eQR code and, after the scanning of the eQR code, to reconstruct the intermediate representation.
    It is formed by a header named \textit{QR script header}, an optional header based on the used dialect, and by the encoding of the executable code.

    \item \textit{eQR code}: a QR code containing executable code. It can be obtained by encoding a program within a QR Code using QRscript.
    
    \item \textit{High-level language}: a computer programming language (even visual) designed to be easy for humans to read and write. It is easier to use and more abstract than the intermediate representation.
    
    \item \textit{Intermediate representation}: lower-level representation of the high-level language used to standardize and simplify the translation between the latter and the correspondent eQRbytecode.
    
    \item \textit{QRscipt}: a series of rules on how to embed a programming language into a QR code. In particular, it refers to the definition of the binary code that has to be saved inside of an eQR code and the rules used for its decodification.
    
    \item \textit{QRscript header}: part of an eQR code that is independent of the dialect. It is described in Section~\ref{sec:eQRbytecodeHeader}.
    
    \item \textit{Security}: features of the QRscript header aimed at managing \textit{authenticity} and \textit{integrity}, and eventually the \textit{encryption} of the content of the eQR code.
    
    \item \textit{Security profile}: indicates the security mechanisms activated and exploited in the current eQR code.
    
    \item \textit{URL}: used to permit the connection to an external URL if an Internet or local connection is available.

\end{itemize}

\bibliographystyle{IEEEtran}
\bibliography{bibliography}

\begin{thebibliography}{1}
\providecommand{\url}[1]{#1}
\csname url@samestyle\endcsname
\providecommand{\newblock}{\relax}
\providecommand{\bibinfo}[2]{#2}
\providecommand{\BIBentrySTDinterwordspacing}{\spaceskip=0pt\relax}
\providecommand{\BIBentryALTinterwordstretchfactor}{4}
\providecommand{\BIBentryALTinterwordspacing}{\spaceskip=\fontdimen2\font plus
\BIBentryALTinterwordstretchfactor\fontdimen3\font minus
  \fontdimen4\font\relax}
\providecommand{\BIBforeignlanguage}[2]{{%
\expandafter\ifx\csname l@#1\endcsname\relax
\typeout{** WARNING: IEEEtran.bst: No hyphenation pattern has been}%
\typeout{** loaded for the language `#1'. Using the pattern for}%
\typeout{** the default language instead.}%
\else
\language=\csname l@#1\endcsname
\fi
#2}}
\providecommand{\BIBdecl}{\relax}
\BIBdecl

\bibitem{QRscript-spec}
S.~Scanzio, M.~Rosani, M.~Scamuzzi, and G.~Cena, ``{QRscript specification},''
  in \emph{arXiv}, Mar. 2024, pp. 1--13.

\bibitem{QRstandard}
\BIBentryALTinterwordspacing
{ISO Central Secretary}, ``\BIBforeignlanguage{en}{{Information technology —
  Automatic identification and data capture techniques — QR Code bar code
  symbology specification}},'' International Organization for Standardization,
  Geneva, CH, Standard ISO/IEC 18004:2015, 2015. [Online]. Available:
  \url{https://www.iso.org/standard/62021.html}
\BIBentrySTDinterwordspacing

\bibitem{rfc3629}
\BIBentryALTinterwordspacing
F.~Yergeau, ``{UTF-8, a transformation format of ISO 10646},'' RFC 3629, Nov.
  2003. [Online]. Available: \url{https://www.rfc-editor.org/info/rfc3629}
\BIBentrySTDinterwordspacing

\bibitem{9921530}
S.~Scanzio, G.~Cena, and A.~Valenzano, ``{QRscript: Embedding a Programming
  Language in QR codes to support Decision and Management},'' in \emph{2022
  IEEE 27th International Conference on Emerging Technologies and Factory
  Automation (ETFA)}, 2022, pp. 1--8.

\bibitem{QRtree}
S.~Scanzio, M.~Rosani, M.~Scamuzzi, and G.~Cena, ``{QRtree - Decision Tree
  dialect specification of QRscript},'' in \emph{arXiv}, Mar. 2024, pp. 1--32.

\end{thebibliography}

\section*{Appendix}
\appendix
\section{Exponential encoding}
\label{app:A}
The exponential encoding is used in many parts of the QRscript language to permit the encoding of integer numbers of any size, but coding at the same time small numbers with a reduced number of bits. The number of bits for the coding is doubled each time the bits used are not enough. The exponential encoding permits a compact notation for small numbers without limiting the size of the number to be encoded a priori.

More in detail, given an integer number without sign $X$, the exponential encoding occupies an exponentially increasing number of bits in the binary representation of the number.

Let $n$ bit be the initial number of bits used for the exponential encoding. As an example, an exponential encoding over four bits is characterized by $n=4$. If the number $X$ is greater than or equal to $(2^n - 1)$ means that the value is too big to fit in $n$ bits. In particular, the remaining value that cannot be coded in $n$ bits is $X - (2^n - 1)$. In this case, the exponential coding doubles the number of bits to $2 \cdot n$.

This process is repeated, doubling $n$ every time the number of bits is not sufficient to code number $X$ (i.e., $n_i = 2 \cdot n_{i - 1}$). The starting point ($n_0$) can be any positive integer number greater than $0$. Primarily, in this specification document, $4$ bits are used as a starting point.

Some examples of this representation can be found in Tab.~\ref{tab:expInt}.

\begin{table}[H]
    \footnotesize
    \centering
    \begin{tabular}{ c | c | c | c | c | c }
        Value & Sign & Base & $1^{st}$ extension & $2^{nd}$ extension & $3^{rd}$ extension \\
        Integer & $1$ bit & $4$ bit & $4$ bit & $8$ bit & $16$ bit \\
        \hline
        12      &   & \texttt{1100} &      &          &                  \\
        \hline
        14      &   & \texttt{1110} &      &          &                  \\
        \hline
        15      &   & \texttt{1111} & \texttt{0000} &          &                  \\
        \hline
        120     &   & \texttt{1111} & \texttt{1111} & \texttt{01011010} &                  \\
        \hline
        300     &   & \texttt{1111} & \texttt{1111} & \texttt{11111111} & \texttt{0000000000001111} \\
        \hline
    \end{tabular}
    \caption{Examples of exponential encoding of integer numbers without sign}
    \label{tab:expInt}
\end{table}

\printindex
\end{document}